\documentclass[proof]{WileyASNA-v1}

\articletype{Article Type}%

\received{26 April 2016}
\revised{6 June 2016}
\accepted{6 June 2016}

\newcommand{\sufi}{\mbox{al-\d S\={u}f\={\i}} }
\newcommand{\sufiGen}{al-\d S\={u}f\={\i}'s }
\newcommand{\SufiGen}{Al-\d S\={u}f\={\i}'s }
\newcommand{\Sufi}{Al-\d S\={u}f\={\i} }
\newcommand{\ulugh}{Ulugh B\={e}g }
\newcommand{\ulughGen}{Ulugh B\={e}g's }

\begin{document}

\title{Accuracy of magnitudes in pre-telescopic star catalogues}

\author[1]{Philipp Protte}

\author[1]{Susanne M Hoffmann}

%\author[3]{Author Three}

\authormark{AUTHOR ONE \textsc{et al}}

\address[1]{\orgdiv{Physikalisch-Astronomische Fakultät}, \orgname{Friedrich-Schiller-Universität Jena},\orgaddress{\country{Germany}}}

%\address[2]{\orgdiv{Org Division}, \orgname{Org Name}, \orgaddress{\state{State name}, \country{Country name}}}

%\address[3]{\orgdiv{Org Division}, \orgname{Org Name}, \orgaddress{\state{State name}, \country{Country name}}}

\corres{Susanne M Hoffmann, \email{susanne.hoffmann@uni-jena.de}}

%\presentaddress{This is sample for present address text this is sample for present address text}

\abstract{Historical star magnitudes from catalogues by Ptolemy (137 AD), \sufi (964) and Tycho Brahe (1602/27) are converted to the Johnson V-mag scale and compared to modern day values from the HIPPARCOS catalogue.
The deviations (or ``errors'') are tested for dependencies on three different observational influences. The relation between historical and modern magnitudes is found to be linear in all three catalogues as it had previously 
been shown for the Almagest data by \citet{Hearnshaw99}. A slight dependency on the colour index (B-V) is shown throughout the data sets and \sufiGen 
as well as Brahe's data also give fainter values for stars of lower culmination height (indicating extinction). In all three catalogues, a star's estimated magnitude is influenced by the brightness of its immediate surroundings.
After correction for the three effects, the remaining variance within the magnitude errors can be considered as approximate accuracy of the pre-telescopic magnitude estimates.
The final converted and corrected magnitudes are available via the Vizier catalogue access tool \citep{VIZIER}.}


\jnlcitation{\cname{%
\author{Protte, P.}, 
\author{Hoffmann, S.M.}},  (\cyear{2020}), 
\ctitle{Accuracy of magnitudes in pre-telescopic star catalogues}, \cjournal{AN}, \cvol{}.}

%%\fundingInfo{Funding info text.}

\maketitle

%\footnotetext{\textbf{Abbreviations:} ANA, anti-nuclear antibodies; APC, antigen-presenting cells; IRF, interferon regulatory factor}

\section{Introduction}\label{sec1}
  Many processes in astronomy have long timescales, especially questions on the evolution of stars (just recently, the timescale of Betelgeuse's supernova was publicly discussed). 
  Further examples include close binary systems such as cataclysmic variables (CVs) and their nova eruption behaviour \citep{shara2017_nov1437,vogt2019,hoffmannVogtProtte}, 
  or even supernovae in close binary systems which have the potential to eject runaway stars \citep{NGH2019}. 
  All these questions on the evolution of astronomical objects require long-term observations but our telescopic surveys only reach back for a few decades (in cases of CVs) 
  or roughly two centuries (in cases of sunspot observations \citep{NN2016,neuhaeuser2018}).\footnote{Sporadic telescopic observations have, 
  of course, existed for a bit longer but \textit{systematic surveys} have not been common practice from the early beginning on.} 
  Aiming for conclusions on long-term evolution it is, thus, desirable to include data from non-telescopic observations which could possibly provide a much longer baseline: 
  Far Eastern tradition, for instance, recorded transient phenomena (such as novae, supernovae, and comets) more systematically than the Western one. 
  However, one of the biggest questions in these terms is the transformation of any ancient (or old) description of the \textit{brightness} of the phenomenon. 
  Only in very few cases (e.\,g. 437, SN~1572) the historical records mention daylight visibility. 
  In a handful of cases (e.\,g. 1175, 1203, 1596, and 1603 according to \citet{ho}, and \cite[129--146]{xu2000}), 
  the description refers to the brightness giving Mars, Saturn, or bright stars like Capella ($\alpha$~Aur) and Antares ($\alpha$~Sco) as comparison. 
  \\
  Although we can look up the brightnesses of the planets and fixed stars in a modern star catalogue or model their brightness at a certain date with our knowledge on their variability, 
  it appears worthwhile to study the accuracy of such historical estimates. \\
  As commonly known, Argelander in the 19th century defined a clear method to estimate the magnitude of a given object by comparing it to a couple of stars in the vicinity. 
  The method(s) of earlier astronomers to derive the magnitude of a star or transient object are yet unknown. As, therefore, the numbers in historical star catalogues are hardly reproducible, 
  we try to derive a better understanding of their scattering, error bars, and their dependencies. 

  \subsection{Dependencies and open questions} \label{DepO}
    In particular, the visual appearance of a celestial point source depends on many influences, e.\,g. the brightness of the background, 
    the local and temporal conditions of the atmosphere and the constitution of the observer's eye. Currently, we cannot consider the observer's 
    eye and we are not even sure that the author of a text book really observed every data point on his own and that not students or assistants were 
    helping or even taking over the measurement. The interplay of the human eye's lens and the atmospheric conditions cause optical effects such as ($i$) 
    reddening of stars close to the horizon (extinction), ($ii$) blurring of bright stars due to humidity or sandstorm, and ($iii$) the impression of rays, horns, 
    or fuzziness of bright objects (see \ref{appendixVenus}) which might affect the estimate of the magnitude. 
    The background brightness of the sky depends on the density of stars in a particular area, the zodiacal light as well as geophysical influences 
    such as the omnipresent airglow (discovered by \citet{Angstrom1869}) and the presence or absence of meteor showers \citep{Siede59}.
    As this work aims to make historical star catalogues usable for modern research, our goal is to find an algorithm for how to deal with historical magnitudes. 
    We assume that magnitudes for a star catalogue have not been observed only one time by one person but cross checked by the assistants of the historical book author. 
    As even Ptolemy mentions the difficulties (and errors) of observations close to the horizon, we assume that they estimated magnitudes at highest possible altitude for a given star. 
    Therefore, the remaining influences to be considered in Section~3 are the dependency of the human brightness estimation on the following questions:
    \begin{itemize}
    \item Redder stars appear fainter, so how does the colour of the star influence the estimate? Can the historical numbers be improved by application of a colour correction? 
    \item The atmosphere influences the appearance by extinction: Does a correction improve the conversion of historical magnitudes into modern ones?  
    \item Observational bias of the environment: How does the presence of bright stars and a background of many faint stars (e.\,g. in the Milky Way) influence the historical brightness estimation?
    \end{itemize} 
    We analyze these questions in Section \ref{sec2} by using the data of three historical star catalogues introduced in Section \ref{originals} which, of course, had already been analyzed by other scholars before us. 
    Their results are, therefore, summarized in the following subsection. 

  \subsection{Placement among previous works}
    The idea to represent different brightnesses by numbers goes back to Antiquity. Pliny the Elder's ($+1$st century) words suggest that Hipparchus ($-2$nd century) might have measured these magnitudes but this 
    cannot be verified or falsified \citep[p.\,92 and 194]{Hoffmann17}. The first surviving appearance of the magnitude scale is in Ptolemy's Almagest from the $2$nd century \citep[p. 4]{HearnBook} and from there it was copied to the Arabic and 
    Latin science culture but the numbers always remained estimations. With the dawning of electric photometry, astrophotography and the necessity of exact values from telescopic observations, 
    the 19th century took some efforts to develop a mathematically exact scale. \citet{Pogson1856}'s system finally prevailed but it differs from the ancient scale because mathematics in the meantime 
    had introduced negative numbers and the zero. Additionally, any type of logarithmic law can only approximate the human sense and it neglects personal influences of the observer (guessing errors). 
    There is no easy conversion from historical magnitudes because the new scale was used for the new star catalogues.     
    \\
    Although the Almagest's star catalogue has thoroughly been analysed and discussed with respect to its record of star positions, there are only a handful \citep[e.g.][]{Hearnshaw99,Schaef13} of recent
    investigations into the star's magnitudes. Even fewer authors consider other pre-telescopic magnitude estimations, which are given most notably by \sufi and Tycho Brahe.
    Recently, \citet{CatBra, CatPU} released online versions of the three catalogues by Ptolemy, \ulugh and Tycho Brahe with \ulughGen catalogue containing magnitude estimations that were 
    adopted from \sufiGen \textit{Book of the fixed stars}.
    \\
    The computer-readable data makes it easier than ever to evaluate and statistically compare the three catalogues.
    If the data optimally converted and corrected for systematic deviations, it might be possible to utilise the ancient magnitudes for investigations of stellar evolution and variability
    (as examples for such endeavours see \citet{Mayer84} and \citet{Hertzog84} although the former's results were later refuted by \citet{Hearnshaw99}).
    All the effects on magnitude estimates, mentioned at the end of Section \ref{DepO} have already been analysed during the second half of the 19th and the first half of the 20th century 
    \citep[for an elaborate example see][]{Zinner26} but there is 
    no systematic query of all three catalogues based on the conversion method, introduced by \citet{Hearnshaw99} while using modern computer-aided statistical procedures.
  
\section{On the original catalogues}\label{originals}
\begin{figure}[h]
    \centerline{\includegraphics[width=78mm]{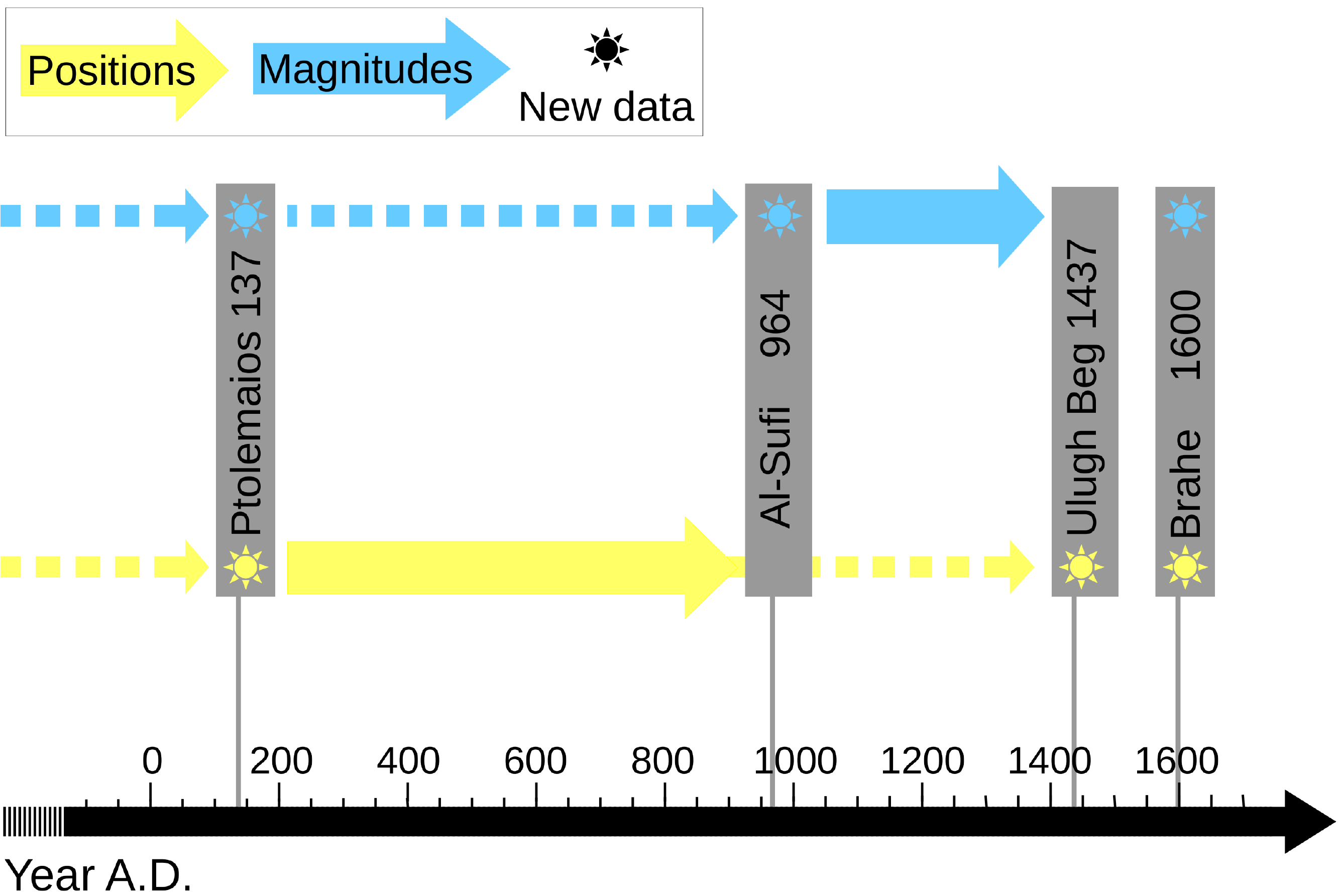}}
    \caption{Transfer of position and magnitude data between the four ancient catalogues. Dashed arrows indicate an uncertain amount of influence, bold arrows show that data was copied almost ``word-for-word''.\label{genesis}}
   \end{figure}

  We evaluated three star catalogues with measurements from different epochs and cultural backgrounds,
   beginning with the one, featured in Ptolemy's $M\alpha\theta\eta\mu\alpha\tau\iota\kappa\acute{\eta}\;\Sigma\acute{\upsilon}\nu\tau\alpha\xi\iota\zeta$
   (engl.: \textit{Mathematical treatise}, commonly known by its Arabic name \textit{Almagest}, 137 AD).
   This work contains the oldest extensive data set of stellar brightness and Ptolemy also was the first astronomer to verifiably make use of a numeric scale: the magnitudes.
   He assigned the brightest stars to the magnitude 1 and the remaining ones into five gradually fainter classes, labelled 2--6.
   For some stars he added qualifiers, saying a star was either slightly brighter or fainter than the given magnitude.
   \\
   The second catalogue is the one by \ulugh from around 1437 AD which contained the first independent, comprehensive position measurements in 1300 years, 
   yet adopted \citep[see][]{Knobel17} its magnitude data from Abd al-Rahman \sufiGen 'Book of fixed Stars' \citep[for a modern english translation see][]{Hafez10},
   which he most likely composed around 964 AD. in the city of Shiraz \citep[p.64]{Hafez10}.
   His list of stars is explicitly based on Ptolemy's catalogue, containing almost the same set of stars with positions, only corrected for precession.
   However, \sufi was only the second astronomer to systematically assign magnitudes to all the entries in his catalogue, using the same numerical scale, as Ptolemy.
   \Sufi's catalogue served as an important source for many subsequent Islamic-Arabic astronomers who used his data or cited his texts \citep[see][p.66 ff]{Hafez10}.
   One of those was \ulugh, who, when he compiled his own star catalogue in 1437, adopted 
   the magnitudes (and in 27 cases also the positions \citep{CatPU2}) from the \textit{Book of the fixed Stars}. 
   \\
   Lastly, we included Tycho Brahe's star catalogue from 1602/1627 which again consists of newly gathered data for positions and magnitudes.
   Brahe was the first modern European scholar to compile a comprehensive original star catalogue. 
   The results were first published as a 777-star catalogue in 1602 shortly after his death but a handwritten copy of a more extensive catalogue had already been sent to several astronomers during the 1590's.
   Finally, in 1627 Johannes Kepler published an edition of the list containing 1004 stars which was very similar to the manuscript version \citep{CatBra}.
   While Tycho set new standards of precision for position measurements, he adopted Ptolemy's magnitude scale with no finer graduation than those of his predecessors.
   As pointed out by most analyses of the different versions of his catalogue \citep[e.g.][]{Baily1843}, the 1627 version even omits the brighter-/fainter-qualifiers that were still included in the previous release.   
   \\
   Figure \ref{genesis} shows the chronology and data transfer of the four catalogues. 

\section{Analysis of the magnitudes}\label{sec2}
   \begin{figure*}[t]
    \centerline{\includegraphics[trim = 1.0cm 7.7cm 1.0cm 7.8cm, clip,width=500pt]{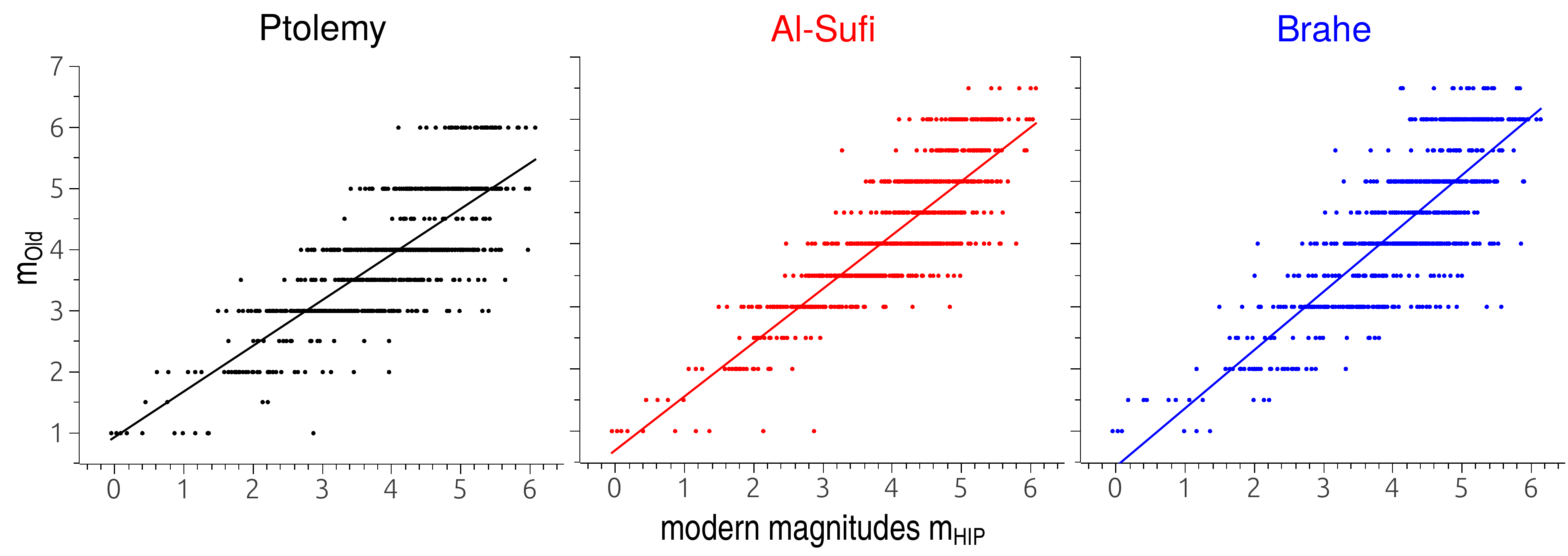}}
    \caption{Relation between historical and modern magnitude scales for each star of the data subsets P, S and B. Although the scatter plot does not look particularly linear, the regression
    can be justified (see text) and is used to convert the historical values to the modern magnitude scale.\label{conv}}
   \end{figure*}
   For the historical magnitudes we used the data, given by \citet{CatPU2,CatBra2}. Additional modern data (e.g. exact position, V-mag and colour index) have been taken from the VizieR release of the HIPPARCOS catalogue \citep{HIP}.
   \\Verbunt and van Gent used the translation of \ulughGen catalogue by \citet{Knobel17} for their digitalisation and the latter, knowing that \ulugh adapted \sufiGen magnitudes, entered them directly from \sufiGen 
   \textit{Book of the fixed Stars} to minimise translation errors. Therefore the catalogue designated as \ulughGen is actually a hybrid catalogue and as only the magnitudes are analysed, they are referred to as 
   '\sufiGen magnitudes' and abbreviated as $m_S$ and the catalogue as $S$.
   \\Brahes catalogue was released in different editions, where only his 777-star list from 1602 includes qualifiers to specify magnitudes beyond 1-mag steps.
   The data given by Verbunt and van Gent includes a total of 1007 stars, merged from all available editions and the qualifiers were included in our analysis whenever they were given.
   \\For the analysis we excluded double entries as well as stars without safe modern identification or magnitude (e.g. for double stars). Furthermore, stars designated as ``faint'' or ``nebulous'' were left out
   and the two brightest stars (Sirius and Canopus) were excluded from most analyses as significant ``outliers``. The used data sets are summarised in Table \ref{data}, where the last set (PSB) only includes stars that were existent in all three
   catalogues and had concordant modern identifications. This set was used to examine the covariance between the historical catalogues.
   Different magnitude designations are used throughout the analysis: $m_\textrm{HIP}$ are the modern V-mag values taken from the HIPPARCOS catalogue, while $m_{old}$ and $m$ designate the historical 
   magnitudes before and after conversion, respectively. Additional indices (P,S,B) might be added when talking about data from a certain author. Finally, we define $\delta m=m-m_\textrm{HIP}$.
   \begin{center}
   \begin{table}[htbp!]%
    \centering
    \caption{The analysed catalogues. Identifications taken from \citet{CatPU2,CatBra2}. PSB is the intersection of P,S,B, containing only stars that are concordantly identified in all three catalogues. \label{data}}%
    \tabcolsep=0pt%
    \begin{tabular*}{20pc}{@{\extracolsep\fill}lcc@{\extracolsep\fill}}
    \toprule
    \textbf{Abbr.} & \textbf{Included authors}  & \textbf{\# of stars} \\
    \midrule
      P 	  & Ptolemy			& 990 		\\
      S 	  & \ulugh			& 988		\\
      B 	  & Brahe			& 937		\\
      PSB	  & Ptolemy+\ulugh+Brahe	& 695		\\
    \bottomrule
    \end{tabular*}
   \end{table}
   \end{center}
    
   \subsection{Conversion of the magnitude scales}
\begin{figure}[h]
	\centerline{\includegraphics[width=0.48\textwidth]{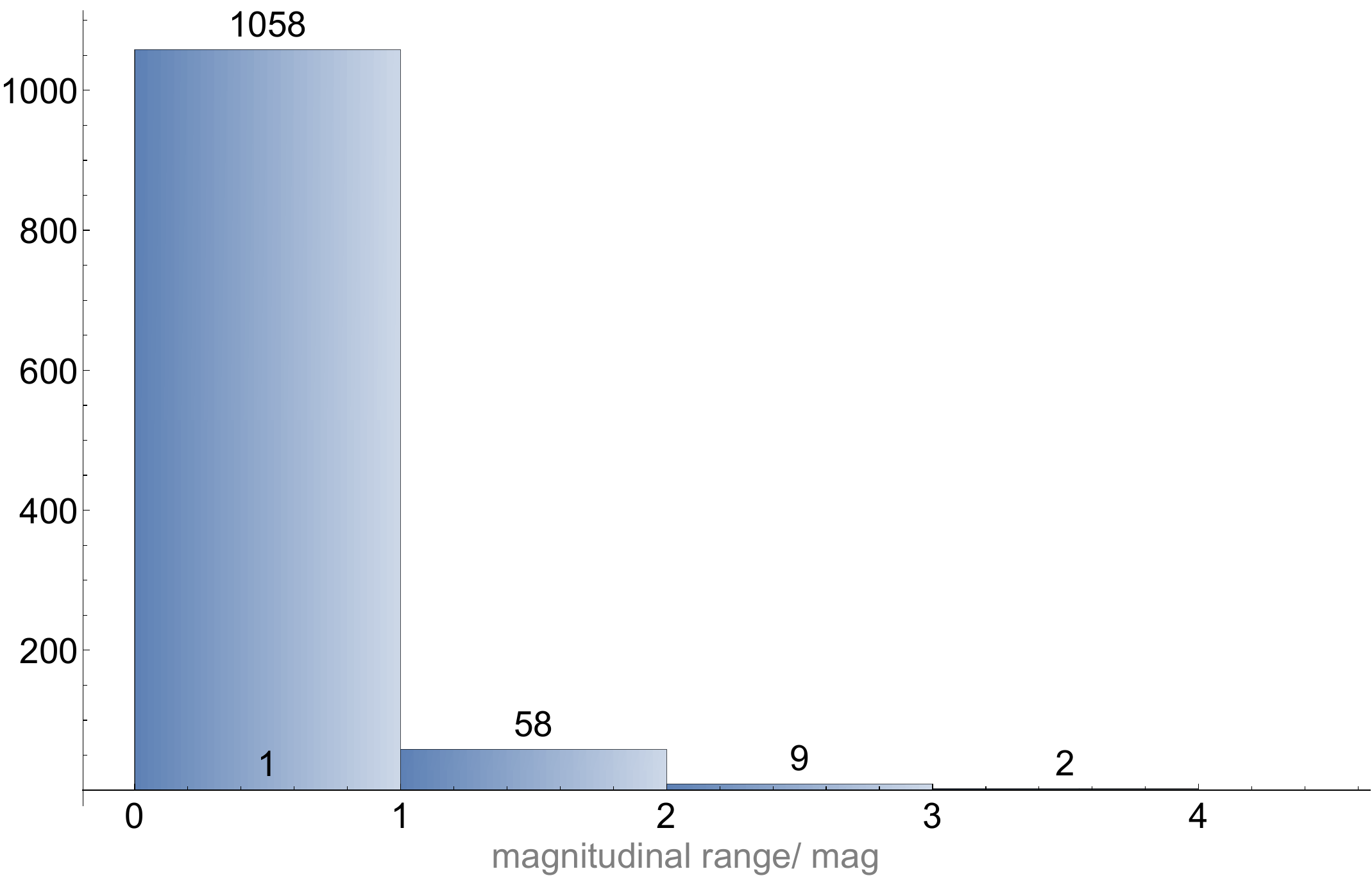}}
	\caption{Histogram showing the distribution of variability amplitudes among the 1,137 known variable stars of $m_{max}, m_{min}<6\,$mag. 
	The possible magnitude error, induced by variability is small, compared to the overall uncertainties of historical magnitudes in most cases.\label{vari}}
      \end{figure}

\begin{figure}[h]
	\centerline{\includegraphics[trim = 3.4cm 4.5cm 2.8cm 4.5cm, clip, width=83mm]{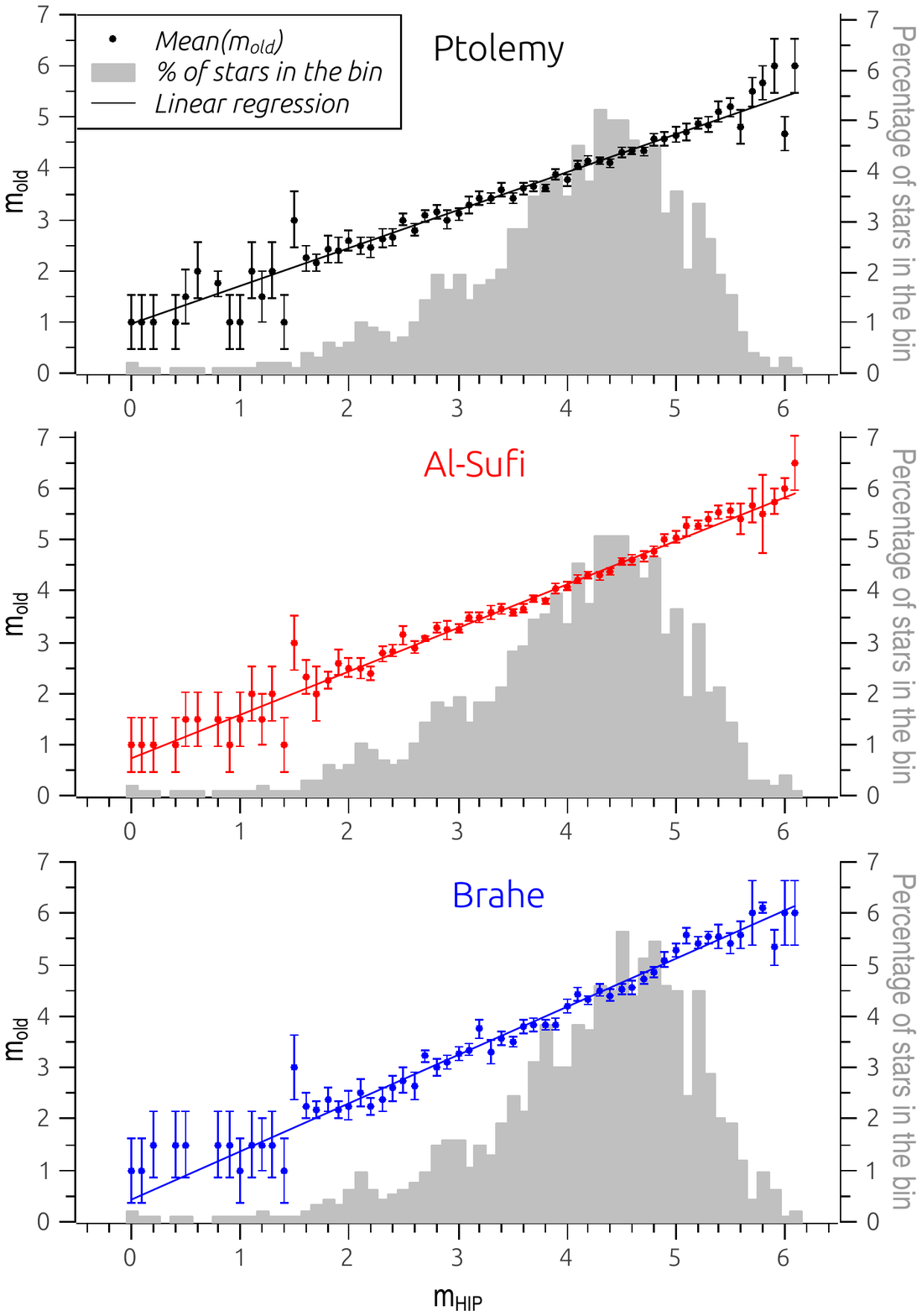}}
	\caption{Average historical magnitudes for modern-magnitude-bins of $0.1$ mag in the data sets P, S, B.
	Error bars are the SEM where bin-size $N>2$ and else the averaged SD of all other bins. 
	The histograms in the background show the relative number of stars within each bin.
	The regression here is only for visualisation of the linearity while the actual coefficients are taken from Figure \ref{conv}.\label{means}}
      \end{figure}

      The historical magnitude values are not identical with modern, photometric V-band magnitudes.
      While today's magnitudes of the stars in our reduced catalogue cover a continuous range from $\sim0$ to $\sim6$,
      in the pre-telescopic era they were based on estimated assignments into 6 discrete groups.
      Nevertheless intermediate steps between these groups were used by all three authors as differently formulated 'qualifiers' which indicate if the star is slightly brighter or fainter than the denoted magnitude.
      \\
      The disparate definitions result in  two problems: on the one hand, the data shows a different range for both magnitudes (roughly $1\dots6 \leftrightarrow 0\dots6$).
      Therefore, a direct comparison of both values in form of a difference $m_{old}-m_\textrm{HIP}$ will be biased towards showing large positive values for brighter stars.
      To minimise any dependency of the difference on a star's brightness, an adequate conversion formula for $m_{old}$ is needed.
      \\
      The other problem, however, has to be tackled first: The intermediate qualifiers, given by the historical observers, do not imply an exact value.
      Trying to convert the qualifiers into numerical divergence, most previous authors added $+0.33$\,mag for the 'fainter-qualifier' and $-0.33$\,mag for the 'brighter-qualifier', 
      but also $\pm 0.3$\,mag and $\pm 0.5$\,mag have been applied.
      In an attempt to find the best approximation, we compared the 'two-step-system' ($\pm 0.33$\,mag) with the 'one-step-system' ($\pm 0.5$\,mag).
      \\
      Looking at the average modern magnitude of each group of stars with qualifiers (e.g. 2(f) -- stars a little fainter than 2nd mag, or 3(b) -- stars a little brighter than 3rd mag), 
      the 'two-step-system' is found to show several inconsistencies. For example, the Almagest's 2(f)-stars which would be identified with $m_{old}^P=2.33\,$mag are fainter on average than the 3(b)-stars, 
      identified with $m_{old}^P=2.67\,$mag.
      The 'one-step-system' on the other hand, is consistent in almost all cases and is therefore adopted for our analysis. 
      The applied values of $m_{old}$ and $m_\textrm{HIP}$ are shown for each star in the scatter plots of Figure \ref{conv}.
      \\
      It might be added at this point that most of the following analysis
      was done with $0.33\,$mag-steps before the 'one-step-system' was chosen. The differences in the results were negligible in almost all cases.
      \\\\
      The modern magnitude scale is logarithmic in regard to the light flux, a fact that corresponds to (and historically derives from) the logarithmic perception of brightness in the human eye (Weber-Fechner-Law).
      Therefore the relation between $m_{old}$ and $m_\textrm{HIP}$ should be approximately linear which is not evident in any of the three sub-figures of Figure \ref{conv}, due to the large scattering.
      But even averaging the $m_\textrm{HIP}$ for each step of $m_{old}$, as it has been done in many previous works \citep[most recently][]{Schaef13}, does not yield a linear correlation, but rather implies a curved function. 
      Thus, instead of trying to find a consistent conversion\textit{ formula} from $m_{old}$ to $m$, most authors applied an empirical method:
      The modern averages are immediately used as $m$ for every star within the respective step.
      \\ 
      Contrary to all previous (and some subsequent) studies,\citet{Hearnshaw99} showed that the linear relation can indeed be found in Ptolemy's data when switching the dependent with the independent variable.
      He argues that taking the mean modern values for each historical magnitude is statistically invalid because the variable with larger uncertainties should be averaged.
      A quick look at the brightest\footnote{$m_{max}, m_{min}<6\,$mag} known variable stars shows that their variability (see Figure \ref{vari}) is usually much smaller than
      the resulting uncertainties of the historic magnitudes (see Figure \ref{delM}).
      \noindent Applying the same procedure as \citet{Hearnshaw99}, we found linear relations for $m_{old}^P$ and for $m_{old}^S$ and $m_{old}^B$, as well which can be seen in Figure \ref{means} where $m_{old}$ is averaged for bins of $m_\textrm{HIP}$.
      It is therefore justified to convert the $m_{old}$ to a new variable called $m$ by applying a linear conversion formula:
      \begin{equation} \label{convert}
	m = a\cdot m_{old}+t \hspace{1cm}\text{with}\hspace{1cm} a = \frac{1}{b} \hspace{1cm} t = -\frac{t'}{b}
      \end{equation}
      Where $b$ and $t'$ are the regression coefficients from Figure \ref{conv} :
      \begin{eqnarray} \label{btstrich}
	b_P&= 0.75 \pm 0.02 \hspace{0.7cm} t'_P&=0.92 \pm 0.08  \nonumber\\
	b_S&= 0.86 \pm 0.02 \hspace{0.7cm} t'_S&=0.69 \pm 0.07  \\
	b_B&= 0.93 \pm 0.02 \hspace{0.7cm} t'_B&=0.43 \pm 0.10  \nonumber
      \end{eqnarray}
     From $b_P$ we can calculate $a_P=1.33\begin{smallmatrix}+0.04\\-0.03\end{smallmatrix}$, meaning that one Ptolemian magnitude corresponds to 1.33 modern magnitudes (concordant with $a_P = 1.36$, as found by \citet{Hearnshaw99}).
     In the same way, the values for the other two authors are: $a_S=1.16\begin{smallmatrix}+0.03\\-0.02\end{smallmatrix}$ and $a_B=1.08\begin{smallmatrix}+0.02\\-0.03\end{smallmatrix}$
     After converting the historical magnitudes to the modern scale, we can now define the error variable $\delta m = m-m_\textrm{HIP}$ for each star.\\

   \subsection{Analysis of the magnitude errors}\label{analysis}

      \subsubsection{Variance and covariance}
\begin{figure}[htbp!]
      \centerline{\includegraphics[trim = 1.6cm 11.7cm 1.6cm 11.3cm, clip, width=83mm]{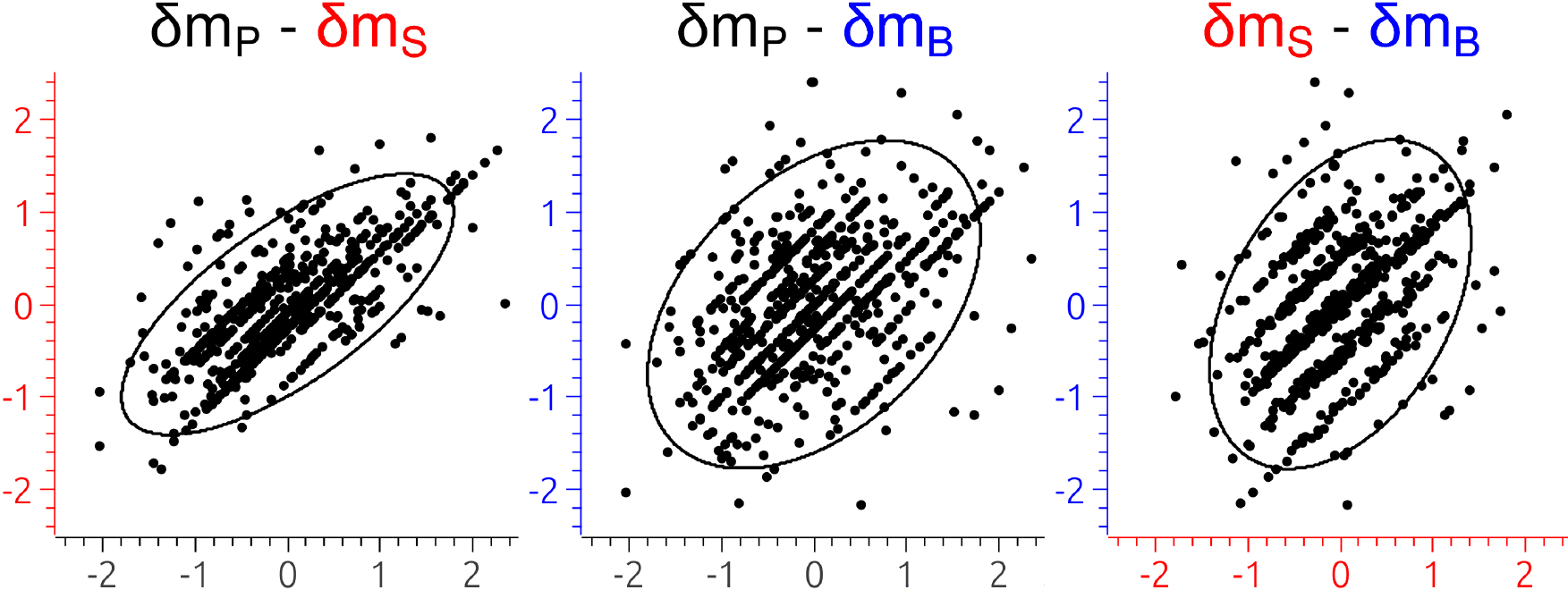}}
      \caption{Correlation between the $\delta m$ of the three catalogues within data set PSB. $2\sigma$-covariance ellipses are given for each scatter plot.\label{Correl}}
      \end{figure}

      In a first step, the distribution of the magnitude errors $\delta m$ is analysed for each of the three star catalogues P, S, B, as well as the common catalogue PSB.
      Table \ref{statist} gives mean values and standard deviations for both cases with the squared correlation coefficients between the $\delta m$ added for the shared data set.
      \begin{table}[htbp] 
      \begin{center}
	\centering
	\caption{Mean values $\mu$ and std.dev. $\sigma$ of the $\delta m$ in the single-author catalogues (P, S, B) and for the shared catalogue (PSB), including squared correlation coefficients $r^2$\label{statist}.}
	\begin{tabular}{c|c|ccc} \toprule
	\multirow{2}{*}{[mag]}	      & \multirow{2}{*}{P, S, B}	&\multicolumn{3}{c}{PSB}	\\	\cline{3-5}
				      &					&$\delta m_P$			&$\delta m_S$			&$\delta m_B$			\\ \midrule
        \multirow{2}{*}{$\delta m_P$ }& $\mu: 0.00$			&$\mu: -0.01$			&				&				\\
				      &$\sigma: 0.79$			&$\sigma: \hphantom{-}0.73$	&				&				\\ \midrule
        \multirow{2}{*}{$\delta m_S$} & $\mu: -0.01$			&\multirow{2}{*}{$r^2: 0.51$}	& $\mu: -0.05$			&				\\
				      &$\sigma: \hphantom{-}0.64$	&				&$\sigma: \hphantom{-}0.58$	&				\\ \midrule
        \multirow{2}{*}{$\delta m_B$} & $\mu: 0.02$			&\multirow{2}{*}{$r^2: 0.18$}	&\multirow{2}{*}{$r^2: 0.16$}	& $\mu: 0.00$			\\
				      &$\sigma: 0.76$			&				&				&$\sigma: 0.72$	\\ \bottomrule        
	\end{tabular}
      \end{center}
      \end{table}
      \noindent The mean values of $\delta m$ for the three single catalogues come close to zero but \sufiGen data in the common list shows a slightly larger offset. 
      This might be due to a selection effect in the shared catalogue where certain stars were omitted.
      The standard deviation shows similar values for Ptolemy's and Brahe's catalogues but a significantly %\footnote{$\Delta\sigma = 2\sigma/\sqrt{2(n-1)}$ is used as certainty of the standard deviation $\sigma$
      %on a 95\% confidence level. This yields $0.04$ mag for P and B, as well as $0.03$ mag for S.} 
      slimmer scattering for \sufiGen magnitude errors. 
      Finally, the $r^2$-values indicate a strong correlation between 
      Ptolemy's and \sufiGen magnitude errors while Brahe's data seems to be largely independent. The correlation is visualised in the scatter plots in Figure \ref{Correl}, including covariance-ellipses that contain $\sim95\%$
      of the data points.
      \noindent The correlation between Ptolemy and \sufi should come as no surprise because \sufi takes Ptolemy's magnitude estimation as a basis for his own (see also \mbox{Figure \ref{genesis})}.
      He even gives literal references like: ''The fourth [star] [$\dots$] is much greater then [sic] 4th magnitude, 
      but it was mentioned by Ptolemy as 4th magnitude exactly.'' \citep[p. 154]{Hafez10}. 
      From these kind of comments it can be assumed that \sufi only changed a star's magnitude if he deemed it distinctly erroneous and therefore 
      left a large fraction of them unchanged, causing this dependency.\\
      The distinctly weaker correlation between Ptolemy's and Brahe's data might as well be due to some of the latter's magnitudes being influenced by the \textit{Almagest}. 
    
      \subsubsection{Dependence on the colour index}
      The transmission spectrum of the modern V-band filter is quite similar but not identical with the spectral sensitivity function of the human eye. In fact there are at least two such sensitivity functions (one for 
      daylight- or \textit{photopic} vision and one for night- or \textit{scotopic} vision) and the V-band curve lies in between of them both.
      Generally, the photopic vision is employed under brighter ambient light and means that the eyes' cones are active and we can perceive colour.
      In contrast, the scotopic vision is the extreme case where our vision depends solely on the eyes' rods which can only differentiate between bright and dark but not detect colour \citep[e.g.][p. 178]{Clauss18}.
      As the rods are more sensitive to shorter wavelengths, the sensitivity function for the scotopic vision is shifted towards blue colours, 
      making blue stars a little brighter than red stars of the same V-band magnitude.
      The upcoming analysis is restricted to stars of $m_\textrm{HIP} > 2.3$, excluding those stars which are bright enough to be seen with photopic vision 
      (indicated by them having a colour to the naked eye)\footnote{The exact threshold of colour vision in terms of star magnitudes is not clearly defined but 
      from observational experience, 2.3\,mag could be an approximate value.}.
      In this case we would expect the calculated $\delta m$ to be something like a colour index between the V-band filter and the human eye-``filter'' for $m > 2.3\,$mag.
      A plot of two colour indices (e.g. ($I-J$) against ($K-L$)) shows an almost linear relation with a slope $\partial_{BV}$ that can be calculated from the effective wavelengths of the four filters \citep{Balle12}.
      We performed a linear regression analysis for plots of $\delta m$ against the ($B-V$)-values for the three catalogues. 
      \\To better visualise the tendency among the data, averages of $\delta m$ were calculated for $(B-V)$-bins of $0.1$\,mag. The result can be seen in the left column of \mbox{Figure \ref{Dep}.} together with a scatter plot of 
      the entire catalogues. A linear model was fitted to the scatter plots and the parameters are shown in Table \ref{paras}.
      From the slopes of the linear regression, it is possible to calculate the effective wavelength of the human eye (under the given premises). The three catalogues yield values from $527\,$nm to $532\,$nm with error bars
      of $\pm 6\,$nm. Both, the increase of $\delta m$ for reddish stars as seen in Figure \ref{Dep}, as well as the calculated effective wavelength falling short of the V-band filter ($\lambda_{eff}=548\,$nm), 
      agree with the above assumption of predominantly scotopic vision. Nevertheless, the calculated wavelength is longer than what would be expected for exclusively scotopic vision \citep[$\sim 507\,$nm, see][]{CVRL}. 
      The calculated regression coefficients make it possible to systematically adjust $m$ with regard to the colour index of each star (see section \ref{formula}).
      Lastly, a similar analysis was attempted for the brighter stars, but given the small number of stars with $m < 2.3\,$mag and the large standard deviation of $\delta m$, 
      the statistical analysis did not yield significant results for any of the three catalogues.
      \begin{table}[htbp] 
      \begin{center}
	\caption{Parameters of the fitted models in Figure \ref{Dep}. For the two linear models $\partial$ is the slope and $\theta$ the intercept.
	$n$ is the number of stars included in each model and $R^2$ the fraction of variance, explained by the model. For statistical testing of the models, see \ref{statistsig}. \label{paras}}
	\begin{tabular}{l|cc|cc|cc} \toprule
			  & \multicolumn{2}{c|}{Colour}	 	& \multicolumn{2}{c|}{Extinction}	&\multicolumn{2}{c}{Background}	\\ \toprule
			  &$ \partial_{BV}$&$ \theta_{BV}$	&\multicolumn{2}{c|}{$k_{fit} $}		&$ \partial_{\beta}$&$ \theta_{\beta}$\\ 
	  P	 	&  0.11&-0.06				&\multicolumn{2}{c|}{-0.01}		&-0.37&-0.69\\ 
	  S		&  0.14&-0.06				&\multicolumn{2}{c|}{0.09}		&-0.41&-0.75\\ 
	  B		&  0.14&-0.06				&\multicolumn{2}{c|}{0.05}		&-0.30&-0.53\\ \midrule
			  &$ n$&$R^2$				&$ n$&$R^2$				&$ n$&$R^2$\\ 
	  P	 	&922 &0.006				&992 &0.0003				&990 &0.024\\ 
	  S		&925 &0.017				&990 &0.022				&989 &0.045\\ 
	  B		&886 &0.011				&938 &0.016				&936 &0.016\\ \bottomrule
	\end{tabular} 
      \end{center}
      \end{table} 
      
      \subsubsection{Extinction features}
      \begin{figure}[htbp!]
	\centerline{\includegraphics[trim = 3.4cm 0.1cm 3.35cm 0.2cm, clip, width=83mm]{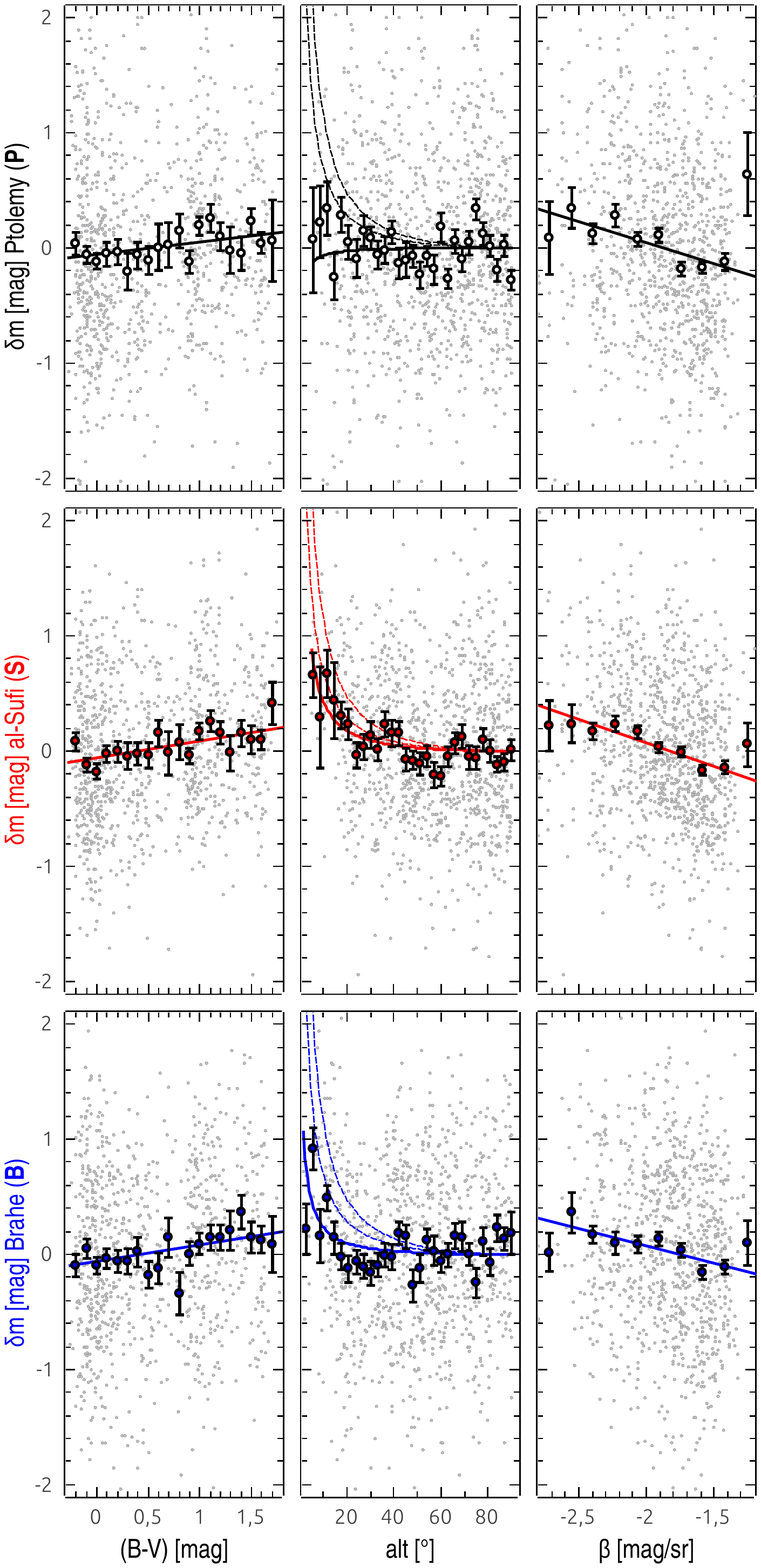}}
	\caption{The dependency of $\delta m$ on colour index \mbox{\textit{B-V}} \textbf{(left)}, maximum culmination altitude \textit{alt} \textbf{(middle)} and 
	background brightness \textit{$\beta$} \textbf{(right)} within catalogues P, S and B.
	Grey scatter plots are the single stars and bold coloured dots are mean values of bins of the independent variable with SEM-error bars.
	The bold lines show models, fitted to the scatter plots for each dependency.
	Stars with $m_\textrm{HIP}<2.3\,$ mag were excluded from the colour-analysis.
	For the middle column, the dashed lines show expected extinction coefficients of $k = 0.15\; \text{and} \;0.25\,$mag/$X$.
	The $\beta$-bins correspond to the levels of grey in \mbox{Figs. \ref{PSBack} and \ref{BBack}.
	All model parameters can be found in Table \ref{paras}.}\label{Dep}}
      \end{figure}
      Even though the effect of atmospheric extinction is obvious in its existence for everyone who has watched the (night-) sky with some attention, no discussion of the effect can be found in any of the works containing the
      three catalogues. If the observers had completely ignored the extinction, it would have to be expected that stars with low culmination altitudes were estimated too faint.
      \citet{Schaef13} analyses the dependency and comes to the conclusion that all three catalogues are in some way ``corrected'' for extinction\footnote{That does not necessarily mean they were explicitly
      corrected by some formula or observational procedure but could also mean to just estimate slightly brighter magnitudes for low standing stars. 
      Trying to observe stars at their highest position could also be considered such a correction and is presupposed for the following analysis.}.
      \\
      With the new conversion, a similar analysis is shown in the middle column of \mbox{Figure \ref{Dep}}, as a scatter plot of all stars, and as averaged $\delta m$-values for $3^\circ$-bins of culmination altitude. 
      Schaefer gives the following extinction function:
      \begin{equation}\label{ext}
	\delta m = k \cdot \left[\left(\sin(alt)+0.025\cdot e^{-11\cdot \sin(alt)}\right)^{-1}-1\right]
      \end{equation}
      With the extinction coefficient $k$, given in magnitudes per airmass $X$ and the horizontal altitude $alt$ of the respective star.
      In application to the historical catalogues, the altitude (or rather the culmination point of a star) can be calculated from the geographic latitude $\phi$ of the observer 
      and the declination $\delta$ of the star at the time of observation.
      \begin{equation}
	alt = 90°-|\delta-\phi|
      \end{equation}
      The extinction curves are plotted within each sub-figure for $k = 0.25\,$mag/$X$ \citep[as suggested by][]{Schaef13},
      as well as $k = 0.15\,$mag/$X$. \citet{Pick02} assumes such a value for a pre-industrial atmosphere.
      Additionally, models according to equation \ref{ext} were fitted to the data in Figure \ref{Dep} with the resulting parameters $k_{fit}$ listed in Table \ref{paras}.
      Schaefer's general result is reproduced with close to no (in fact even a slight but insignificant negative)
      extinction effect showing in the \textit{Almagest's} data. In contrast, an effect is 
      clearly visible in \sufiGen magnitudes but it still falls short of
      the expected intensity for Schaefer's extinction coefficient. However, the data could almost agree with the lower extinction coefficient of $k = 0.15\,$mag/$X$.
      In Brahe's magnitude estimations, a weak extinction effect can be found but again, it falls way short of the plotted models. 
      Although the effect is weaker than expected, it can be corrected for at least in \sufiGen and Brahe's data, as a clear systematic deviation can be found there.
      \\
      It should be noted that the models, fitted in Figure \ref{Dep} are very sensitive to single extreme $\delta m$ at low altitudes,
      which might occur due to falsely identified stars or other sporadic errors. 
      Furthermore, it also seems to be highly controversial what extinction coefficient would have to be expected for a pre-industrial atmosphere \citep[see][]{Hearnshaw99, Pick02, Schaef13}\footnote{Possibly, 
      the worldwide 2020 Corona-lockdown might bring new insights into this question.} and lastly, 
      the actual effect found in each catalogue also depends on the exact method by which the magnitudes were estimated which can only be speculated about.

      \subsubsection{Star maps and background brightness}
      The dependency of $\delta m$ on the colour index, as well as the culmination altitude, are both effects that can be understood and modelled in a (bio-)physical sense.
      However, there seem to be further trends within the data, which can be found looking at the spatial distribution of the $\delta m$.
      \begin{figure*}[htbp]
	\centerline{\includegraphics[width=500pt]{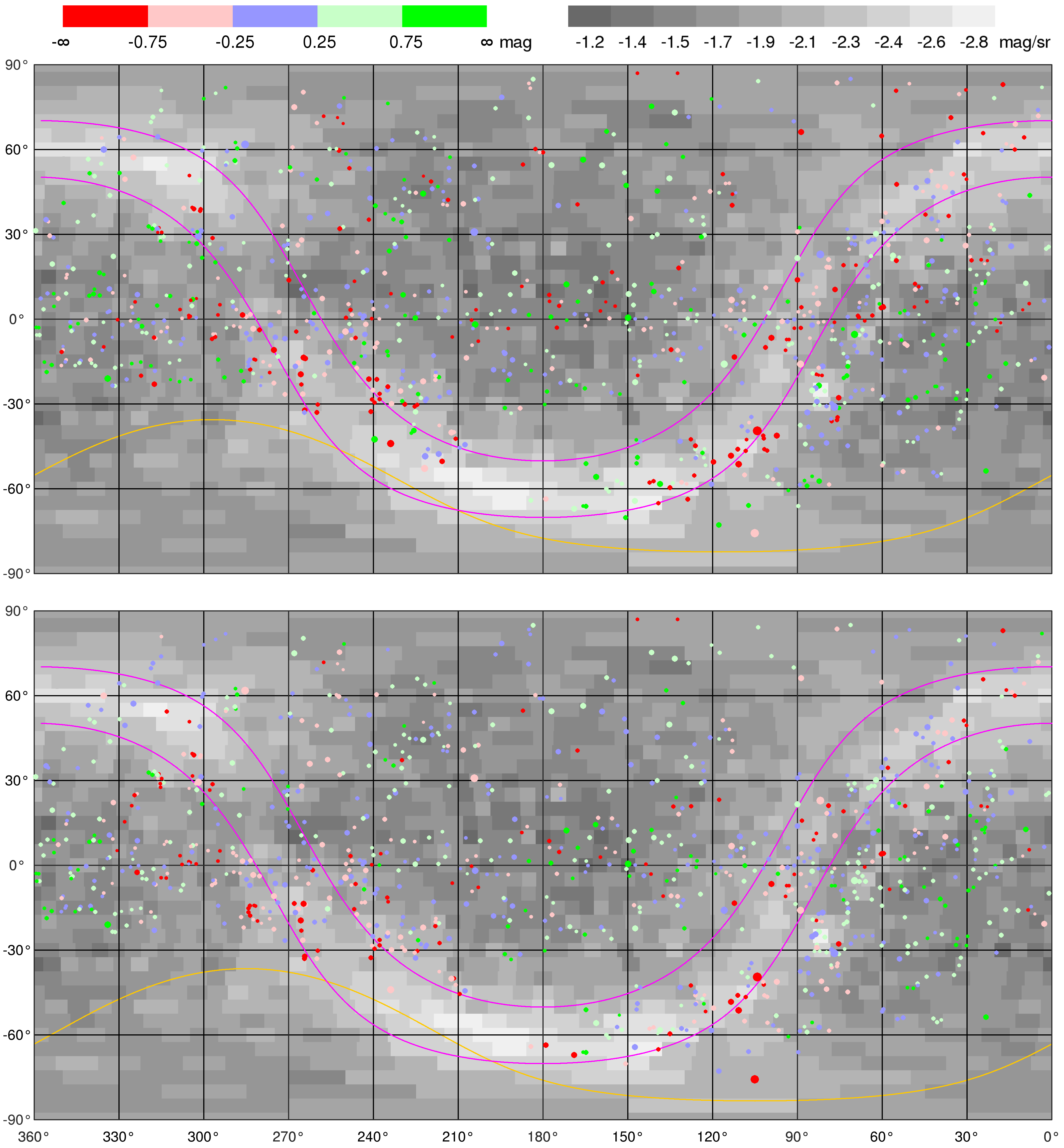}}
	\caption{Plots of star catalogues P (top figure) and S (bottom figure) in ecliptical coordinates (equinox J2000). The stars' size is according to their $m_\textrm{HIP}$ and colour according to their $\delta m$. 
	Green stars were estimated too bright and red ones too faint (see scale). The pink lines mark the Milky Way
	($\pm 10^\circ$ of gal. lat.), the orange lines are the southern visibility limits of the respective time and place. The background depicts the summed brightness of stars in the area as grey-scale.
	More precisely, the flux of all stars between $6\,\text{mag}<m_\textrm{HIP}<10\,\text{mag}$ from the HIPPARCOS catalogue is summed and given as surface brightness of the respective area. 
	For details on background-colouring, see text. There is a clear tendency of 
	many too brightly estimated stars in darker areas and vice versa. \label{PSBack}.}
      \end{figure*}
      \begin{figure*}[htbp]
	\centerline{\includegraphics[width=500pt]{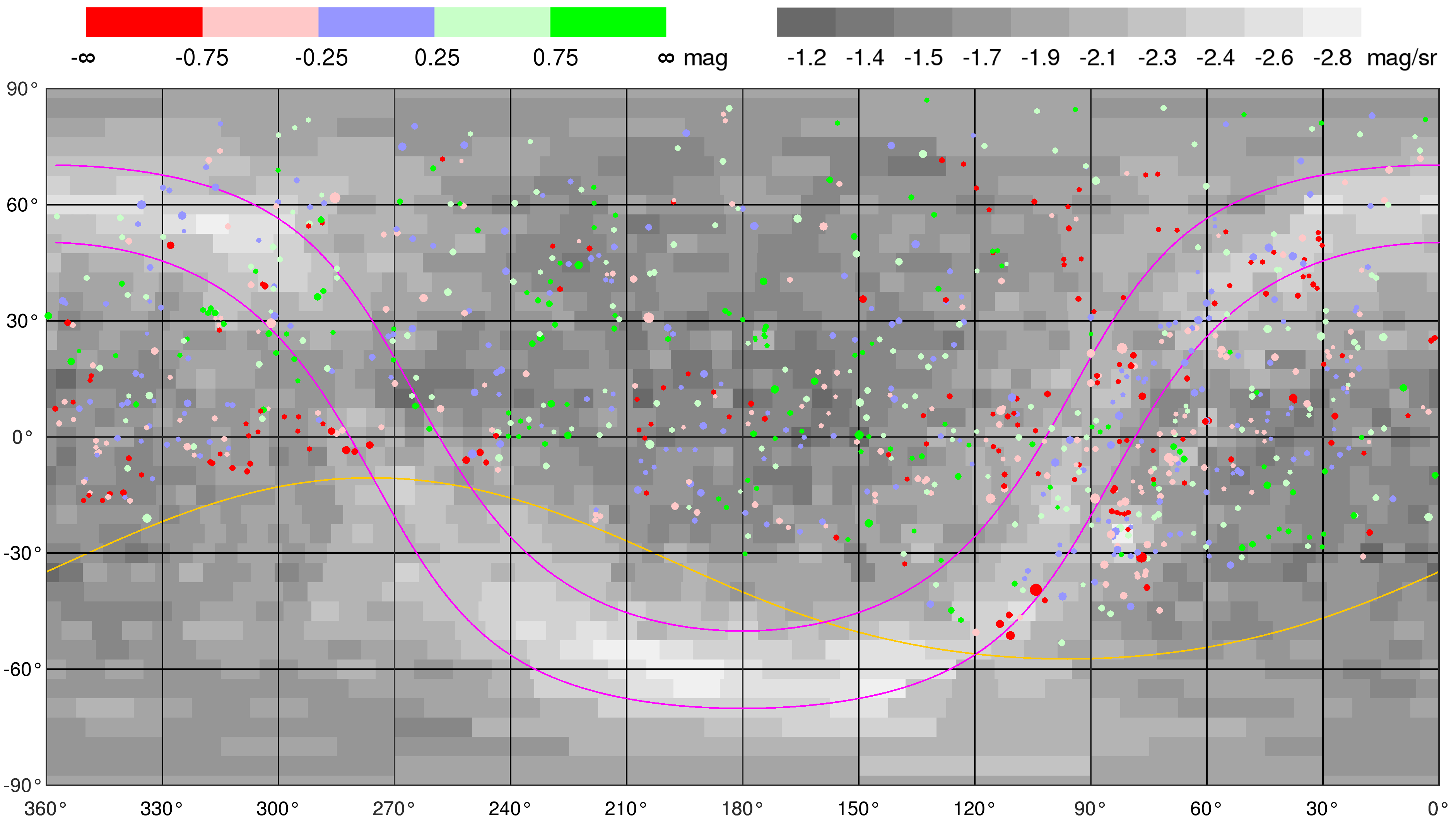}}
	\caption{Plots of star catalogue B in ecliptical coordinates (equinox J2000). For detailed description, see Figure \ref{PSBack} \label{BBack}}
      \end{figure*}
      Figures \ref{PSBack} and \ref{BBack} show maps of all stars within catalogues P,S,B. The $\delta m $-values are rounded into 5 bins and the stars coloured accordingly.
      The exact colour-scales can be found within both figures. Additionally, the maps show a kind of ``background brightness'' or rather a summed brightness of stars per area, which is depicted by a graduated grey-scale.
      The actual area from which the flux is summed are not the grey rectangular fields, but circles -- or rather cones in actual 3D-space -- around the centre of each field.
      These cones have a uniform radius of $4^\circ$ each, so the summed flux can easily be converted to mag/sr. 
      The grey fields are not of perfectly uniform solid angle but were necessary for the visualisation, as they cover the whole projection  without gaps or overlaps. The fields cover $5^\circ$ of latitude each 
      and a longitude segment that corresponds best to $5^\circ$ of a great circle while still guaranteeing an integer amount of segments within the $360^\circ$ circle of latitude.
      For the summation, all stars within the HIPPARCOS catalogue \citep{HIP}, between 6\,mag and 10\,mag were used (a total of 112{,}914 stars). Only faint stars were chosen for several reasons:
      \begin{itemize}
	\item Assure that a star's background is not primarily defined by the star itself.
	\item Keep the differences in surface brightness small, even for a high spatial resolution (i.e. small grey fields).
	\item The density of those dimmer stars corresponds well to the perceived brightness of the actual night sky (e.g. the Milky Way (MW) is clearly visible)
      \end{itemize}
      Looking at the maps, we can find areas within each catalogue where stars are predominantly estimated too faint (``red areas''). 
      More specifically, we can make the following observations for catalogues P an S:
      \begin{enumerate}
	\item The similarity of catalogues P ans S can be found once more in the maps.
	\item Nevertheless, \Sufi seems to have reworked many of Ptolemy's most southern stars to fainter values.
	\item Both show ``red areas'' throughout the MW and especially around the centre of the galaxy ($240<\lambda<300^\circ\,/\,0<\beta<30^\circ$).
	\item Like Ptolemy, \Sufi still estimates many stars in the area $300<\lambda<360^\circ\,/\,-30<\beta<30^\circ$ as too bright. 
	  However the visibility limit shifts away from those stars and towards the galaxy centre for \sufiGen time.
	\item \Sufi generally has fewer extreme mistakes, depicting his lower standard deviation in $\delta m$.
      \end{enumerate}
      Points 2 and 4 mostly explain the stronger extinction effect in \sufiGen data (see Figure \ref{Dep}) but can only partly (2.) be considered to be really caused by extinction.
      Apart from that, it seems obvious that both authors show a tendency to estimate stars in bright areas too faint.
      \\
      Brahe's map differs considerably from P and S, showing the following notable features:
      \begin{enumerate}
	\item The brightest parts of the MW are missing, due to Brahe's northern geographic latitude.
	\item ``Red areas'' can be found from the galaxy centre (only partly visible) along the visibility limit (VL) within $270<\lambda<330^\circ$.
	\item However, other areas along the VL are not particularly red.
	\item Another large ``red area'' can be found at $30<\lambda<120^\circ$, roughly along the MW.
	\item Again, other parts of the MW do not show any clear tendency.
      \end{enumerate}
      It becomes obvious, where the extinction feature (see middle column of Figure \ref{Dep}) stems from in Brahe's case (2.) but it is remarkable that the effect is obvious within this area and
      completely vanishes for other longitudes.
      A dependency on the background brightness can also be found in some parts of the map.
      For all three catalogues it is very notable that effects of background brightness and extinction are partly visible but can not be the sole reason of every ``red'' or ``green area''.
      \\
      Concerning the dependency of $\delta m$ on the background brightness, one might want to explain it by the varying degree of dark-adaptions for differently
      bright areas. However, single bright stars would have the strongest effect here and those are excluded from the background brightness, shown in the maps
      \footnote{looking at the surroundings of the brightest stars, there is no clear trend in any of the maps, either way.}.
      So the whole phenomenon seems to be of rather psychological nature which makes it harder to quantify theoretically.
      Nevertheless, several authors \citep{Zinner26, Hearnshaw99} have described and analysed the dependency but mostly restricted themselves to a comparison between stars within and outside of the Milky Way.
      Going a step further, we used the value of background brightness from the maps (Figs. \ref{PSBack},\ref{BBack}) to plot average $\delta m$ for 10 bins of surface brightness. 
      The mean values of each bin are shown in the right column of Figure \ref{Dep} together with a scatter plot of all stars. The decline of $\delta m$ for darker backgrounds becomes clearly visible.
      As there is no available mathematical model to describe the expected dependency of $\delta m$ on the background brightness, a linear regression is the simplest approximation.
      The regression parameters (see Table \ref{paras}) can again be used to correct the values of $\delta m$ for the described effect.
      \begin{table}[htbp] 
      \begin{center}
	\caption{Average $\delta m$ of stars within ($\delta m_{MW}$) and without ($\delta m_{notMW}$) the Milky Way.\label{MW}}
	\begin{tabular}{l|cc} \toprule
	  Cat.     	& $\delta m_{MW}$		& $\delta m_{notMW}$\\	\midrule
	  P	 	&$ 0.23	\pm 0.06$		&$-0.06 \pm 0.03$	\\ 
	  S		&$ 0.24 \pm 0.05$		&$-0.07 \pm 0.02$	\\ 
	  B		&$0.08 \pm 0.06	$		&$0.01 \pm 0.03$	\\ \bottomrule
	\end{tabular} 
      \end{center}
      \end{table}
      Of course, the correction can only be made if the exact same background brightness values are calculated for every star which might be laborious for anyone trying to make use of the correction formula \eqref{corr1}.
      As an alternative, the average $\delta m$ for stars within and without the Milky Way ($\pm 10^\circ$ of galactic lat.) is given in Table \ref{MW} and 
      can be used for the correction formula \eqref{corr1} instead.
    
\section{Conclusions}\label{conc}
  \begin{figure*}[t]
    \centerline{\includegraphics[trim = 2.2cm 5.3cm 2.1cm 5.5cm, clip,width=500pt]{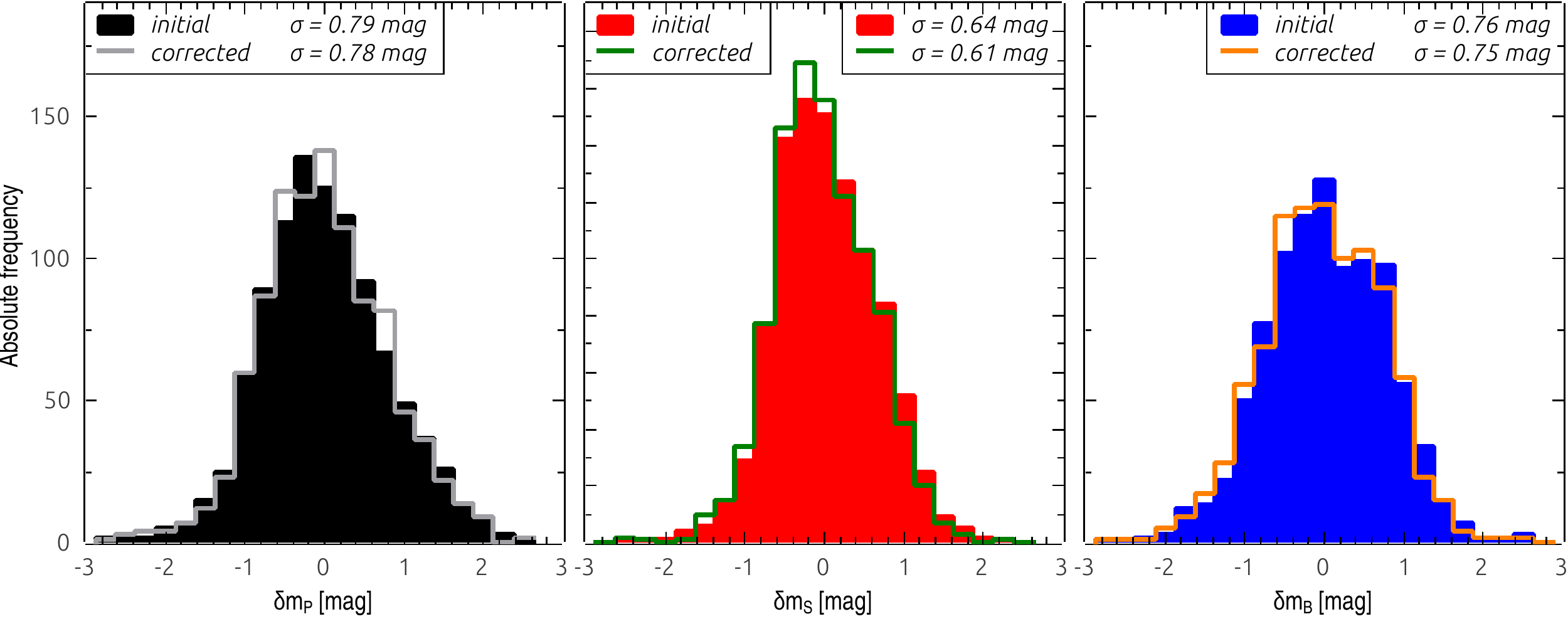}}
    \caption{Distribution of the $\delta m$ before and after the corrections, applied in section \ref{analysis}. The histograms of bin size $0.25\,$mag show approximate normal distributions 
    for all three catalogues. The corrections result in only minor changes of the distributions. Overall, \SufiGen magnitude errors show the lowest variance. \label{delM}}
  \end{figure*}
  The above analysis investigated different systematical dependencies within the magnitude data of the three historical catalogues.
  It was shown that stars are often estimated slightly fainter, if \dots
  \begin{itemize}
    \item they are red 
    \item they stand close to the southern visibility limit 
    \item they are seen in bright areas of the night sky.
  \end{itemize}

  \subsection{Correction formula} \label{formula}
    The effects were described quantitatively and can therefore be corrected. 
    As the single corrective terms are small in comparison to the overall variance in $\delta m$, they can be assumed to be independent from one another and the correction takes the form of a simple additive parameter
    $\Delta_i$ for each of the three models which is subtracted from the initial converted magnitude.
    This yields a new magnitude $m^*$ for each star which can be considered the best approximation to the modern V-mag scale.
    \begin{equation} \label{corr1}
      m^* = \frac{m_{old}-t'}{b} -\Delta_{B-V} -\Delta_{Ext} -\Delta_{\beta}
    \end{equation}
    Where $t'$ and $b$ are given in equation \eqref{btstrich}.
    The single $\Delta_i$ are then the respective models which were fitted to the data, or for the background brightness it can also be the alternative model based on the position within or without the Milky Way.
    \begin{eqnarray}
      &\Delta_{B-V} &= \partial_{B-V} \cdot (B-V) + \theta_{B-V} \nonumber\\
      &\Delta_{Ext} &= k \cdot\left[ (\sin(alt)+0.025\cdot e^{-11\cdot \sin(alt)})^{-1}-1\right]  \nonumber\\
      &\Delta_{\beta} &= \partial_{\beta} \cdot \beta + \theta_{\beta} \hspace{0.5cm} \text{\textbf{\textit{or}}}\hspace{0.5cm} =\delta m_{MW/notMW} \nonumber
    \end{eqnarray}
    The empirical parameters $\partial_{B-V}, \theta_{B-V}, k, \partial_{\beta}, \theta_{\beta}$ can be taken from Figure \ref{Dep} and $\delta m_{MW/notMW}$ is given in Table \ref{MW}.
    
  \subsection{Statistical significance} \label{statistsig} 
    Looking at the three single correction models, it is conspicuous that the variance, explained by the models is much lower than the residual variance induced by the wide scattering of the $\delta m$.
    In fact the $R^2$-values (see Table \ref{paras}) which express the fraction of the variance, explained by the models, remain at a few percent for all nine models.
    However, when testing the models for significance against a null hypothesis $H_0$ which predicts zero correlation, $H_0$ can be dismissed (meaning $F_n>F_{1,n-2,0.99}$) in almost all cases on a $95\,\%$ or higher
    confidence level. This is due to the high number of $\delta m$ values and suggests that the corrections -- however little variance they might explain -- are in fact significant.
    The only model, not showing statistical significance is the extinction correction for the \textit{Almagest} data, which showed a negative (and therefore nonsensical) value of $k_{fit}$.
    \begin{table}[htbp] 
    \begin{center}
      \caption{The models in Figure \ref{Dep} were tested for statistical significance against the hypothesis $H_0$ of no dependency.
      The resulting parameters $F_n$ were calculated from $R^2$ and $n$ (see Table \ref{paras}) and are given together with the confidence level (CL) on which $H_0$ can be rejected.
      \label{Ftest}}
      \begin{tabular}{l|cc|cc|cc} \toprule
			& \multicolumn{2}{c|}{Color}		& \multicolumn{2}{c|}{Extinction}	&\multicolumn{2}{c|}{Background}	\\	
			& $F_n$		&CL			&  $F_n$	&CL			&  $F_n$		&CL	\\ \midrule
	P	 	&$ 5.6$		&$95\% $		&$ 0.3$		&--			&$ 23.9$		&$99.9\% $	\\ 
	S		&$ 16.0 $	&$99.9\% $		&$ 22.2$	&$99.9\% $		&$ 46.5$		&$99.9\% $	\\ 
        B		&$ 9.8 $	&$99\% $		&$ 15.4$	&$99.9\% $		&$ 15.0$		&$99.9\% $	\\\bottomrule
        \multicolumn{7}{l}{$F_{1,n-2,0.95/0.99/0.999}$ = 3.9 / 6.7 / 10.9 for $886<n<992$} 
      \end{tabular} 
    \end{center}
    \end{table} 
    Nevertheless, when applying the corrections, the low values of $R^2$ lead to an almost negligible ``improvement'' of the $\delta m$'s standard deviations. 
    Figure \ref{delM} shows the distribution of the $\delta m$ before and after the correction formula was applied.
    
  \subsection{Final Remarks on the accuracy of pre-telescopic magnitudes}
    So after all, what is the accuracy of pre-telescopic magnitude estimations and how can they best be converted to their corresponding V-mag values? 
    From the three major catalogues which contain original magnitudes, Ptolemy's Almagest and Tycho Brahe's data can be considered largely independent and both show uncertainties of very similar size.
    \SufiGen estimations, on the other hand can be seen as an -- in most instances -- improved version of Ptolemy's data which show a significantly higher accuracy.
    The distribution of magnitude errors $\delta m$ in all three catalogues follows almost Gaussian curves, so the doubled standard deviations $2\sigma$ can be considered error bars on a 95\% confidence level.
    \\
    Any magnitude, taken directly from one of the catalogues to be used for studies of transient observations and longterm variabilities or even processes of stellar evolution, 
    should first be converted before comparing them to modern V-mag values.
    We recommend, adding the brighter- / fainter-qualifiers as $0.5\,$mag-steps to the original magnitudes and then employing formula \eqref{convert} to attain the Johnson V-magnitude.
    The resulting values should be sufficient for most applications and come with error bars of:
    \begin{equation*}
      2\sigma_{P} = 1.59\, \text{mag}\;,\;\; 2\sigma_{S}= 1.26\, \text{mag}\;,\;\; 2\sigma_{B}= 1.52\ \text{mag}
    \end{equation*}
    In comparison, \citet{Hearnshaw99} calculates standard deviations between 0.41 and 0.72$\,$mag for most groups of Ptolemian magnitudes while \citet{Zinner26} gives values between 0.44 and 0.60$\,$mag as ``mean errors''\footnote{probably
    mean \textit{absolute} errors which are always smaller than (or equal to) the standard deviation. Both values are given for the historical magnitudes $m_P$ where our own standard deviation results in
    $\sigma'_P=0.79\,\text{mag}/1.33=0.59\,$mag and therefore in good concordance with the previous studies.\\
    In some cases it might be sensible }to adopt the corrected values, even though the error bars are not distinctly reduced by the correction.
    Those magnitudes can either be attained by equation \eqref{corr1} or taken from the online catalogue, provided by the authors (see \ref{dataonline})
    Using the corrected values seems especially necessary when analysing only certain groups of stars which might otherwise be systematically biased. This could for example be red giants which all show high values of (B-V)
    or stars within a certain constellation which might all be in an especially bright or dark part of the night sky.
    \\
    Other than that, the analysis of dependencies on colour, extinction and background brightness might also be used to investigate otherwise unrelated questions like the extinction coefficient of the 
    pre-industrial atmosphere or even the effective absorption wavelength of the human eye (under naked eye observation conditions).
    \\
    As we conclude that the error bars of the magnitudes in historical catalogues are $\sim1.3$ to $\sim1.6$~mag, almost all (> 93 \%) variabilities of the naked eye stars (as displayed in Fig.~\ref{vari}) 
    are covered by the error bars which makes it virtually impossible to detect longterm variabilities
    \\
    It should, however, be kept in mind that statistics mean little for a single data point.
    As we know, for particular cases, the ancient observers must have recognised changes in brightness of less than $1.3\,$mag.
    After all, it was possible to naked eye observe the brightness drop of Betelgeuse ($\alpha$ Ori) in winter 2019/ 2020 for many laymen
    \footnote{Also, Aboriginal Australians seem to have discovered Betelgeuses' variability \citep{Schaefer18}} and there are hypotheses 
    that the variability of Algol ($\beta$ Per) had been known in ancient Egypt \citep{Jetsu13}. 
    Cases like these are possible for individual stars which are in  a region with appropriate naked eye comparison stars. 
    That is why, our statistical error bars should be considered the general first step but for some handpicked individual stars careful case studies appear worthwhile.

\backmatter

\section*{Acknowledgments}
P.P. thanks Prof. Bradley Schaefer, Dr. Rob van Gent and Prof F. Richard Stephenson for answering my requests and helping me get a first grip of the topic during my Master Thesis.\\
We furthermore thank our referee, Prof. John Hearnshaw, for his recommendations and reassuring comments.\\
S.H. thanks the Free State of Thuringia for the employment at the Friedrich Schiller University of Jena, Germany. \\
We thank Ralph Neuhäuser (AIU, Friedrich-Schiller-Universität Jena) for the initiative of investigating historical data and to create room for transdisciplinary projects. \\
This research has thankfully made use of VizieR catalogue access tool, CDS, Strasbourg, France (DOI:10.26093/cds/vizier, \citep{VIZIER}),
and the VSX variable star catalogue of the American Association of Variable star Observers (AAVSO) \citep{AAVSO}. 

\subsection*{Author contributions}
This analysis of pre-telescopic magnitudes was mostly written by PP in the context of his master thesis.
SH developed the idea from the thesis towards this contribution and offered extensive support, 
advice and revision during the completion of the thesis as well as the article.
Additionally, SH contributed large parts of the introduction and the final remarks.

%\subsection*{Financial disclosure}

%None reported.

%\subsection*{Conflict of interest}

%The authors declare no potential conflict of interests.

%\section*{Supporting information}

\appendix
\section{Optical influences of human vision on magnitude estimates}\label{appendixVenus}
\begin{figure*}[ht!]
    \includegraphics[width=\textwidth]{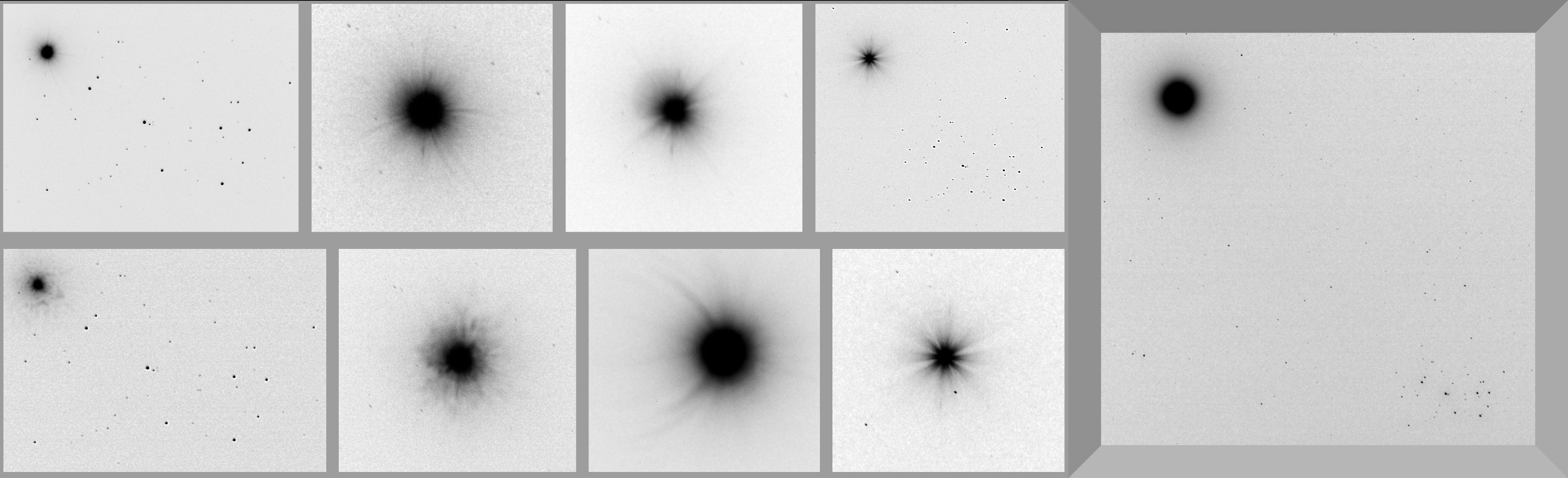}
    \caption{Naked eye appearance of a bright object in contrast to fainter stars (here: Venus, $-4.6$~mag, next to the central part of the cluster of the Pleiades, 
    stars of 2.9 to 4.3~mag). Photos are shown as taken with Canon 600D and 300~mm tele objective, without any astronomical preparation, without any filter or any processing 
    in order to display the real impression for the human eye.\label{venus}}
\end{figure*}

The pictures in figure \ref{venus} show that a bright object (in this case, Venus) can be described as `having rays', `horned', `hairy', or `fuzzy'. 
The photos were taken in central Europe (April 4th to 6th 2020) under normal \textit{clear} weather conditions.
\\
The rays and horns of bright objects are not only an effect of the weather but are produced by the interplay of a lens (of the eye as well as of camera optics) and its entrance pupil with an entering wavefront. 
Passing through a lens with a limited entrance pupil, 
the wavefronts are described by the Zernike polynomials $Z$ producing the known effects like astigmatism $Z_2$, coma $Z_3$, the trefoil effect $Z_3$ (three rays) , 
spherical aberration $Z_4$, and higher orders of aberration in the perfectly spherical lens. 
The effect is caused by the limited size of the pupil \citep{lopezGil2007} and unevenness of the border increases the effect, as well as astigmatism of the lens itself. 
The irises of both, camera and eye, are limitations of the pupil and the polygonal shape of the mechanical iris of a camera lens as well as the muscles at the edge of the eye both increase such effects: 
The photos of these rays do in fact show roughly the same as what the eye sees.
\\
With atmospheric conditions of the desert or in tropical climate (with sandstorm or humidity) the atmospheric effects become stronger and the bright point source is blurred; 
the beam of light from the star does not enter the pupil parallelly and the contraints for applying the Zernike polynomials directly are not fullfilled perfectly anymore. 
The atmospheric influence can even lead to less rays, simply showing the blurred Airy disks around the bright object instead (rightmost picture, with cirrus clouds on April 10th).
    
\section{Online-Only Material} \label{dataonline}
We prepared the catalogues according to our suggestions in the above work. The data files will be uploaded in CDS as soon as the paper is published. 
It will be available at CDS via anonymous ftp to cdsarc.u-strasbg.fr (130.79.128.5)
or via http://cdsarc.u-strasbg.fr/viz-bin/qcat?J/AN

\nocite{*}% Show all bib entries - both cited and uncited; comment this line to view only cited bib entries;
\bibliography{Wiley-ASNA}%

%\section*{Author Biography}
%(if applicable)

%\begin{biography}{\includegraphics[width=60pt,height=70pt,draft]{empty}}{\textbf{Author Name.} This is sample author biography text this is sample author biography text this is sample author biography text this is sample author biography text this is sample author biography text this is sample author biography text this is sample author biography text this is sample author biography text this is sample author biography text .}
%\end{biography}

\end{document}